\documentclass{ws-procs975x65}

\def\PRC{{\em Phys. Rev.} C}
\def\NPA{{\em Nucl. Phys.} A}
\def\AP{\em Ann. Phys. (N.Y.)}

\begin{document}

\title{Everything you always wanted to know about SUSY, 
but were afraid to ask}

\author{R. Bijker$^1$, J. Barea$^1$ and A. Frank$^{1,2}$}

\address{$^1$ ICN-UNAM, AP 70-543, 04510 M\'exico, DF, M\'exico\\
$^2$ CCF-UNAM, AP 139-B, 62251 Cuernavaca, Morelos, M\'exico}


\maketitle

\abstracts{New experimental tests of nuclear supersymmetry are suggested. 
They involve the measurement of one- and two-nucleon transfer reactions 
between nuclei that belong to the same supermultiplet. These reactions 
provide a direct test of the `fermionic' sector, i.e. of the operators 
that change a boson into a fermion or vice versa. We present some theoretical 
predictions for the supersymmetric quartet of nuclei: $^{194}$Pt, 
$^{195}$Pt, $^{195}$Au and $^{196}$Au.}

\section{Introduction}

Dynamical supersymmetries were introduced \cite{FI} in nuclear physics in 
1980 by Franco Iachello in the context of the Interacting Boson Model (IBM) 
\cite{IBM} and its extensions.  
The spectroscopy of atomic nuclei is characterized by the interplay
between collective (bosonic) and single-particle (fermionic) degrees of
freedom. The IBM describes collective excitations in even-even nuclei in 
terms of a system of interacting monopole and quadrupole bosons with angular 
momentum $l=0,2$. The bosons are associated with the number of 
correlated proton and neutron pairs, and hence the number of bosons $N$ is 
half the number of valence nucleons. 

For odd-mass nuclei the IBM has been extended to include single-particle
degrees of freedom \cite{IBFM}. The Interacting Boson-Fermion Model (IBFM)
has as its building blocks $N$ bosons with $l=0,2$ and $M=1$ fermion with 
$j=j_1,j_2,\dots$. The IBM and IBFM can be unified into a superalgebra 
$U(n/m)$, where $n=\sum_l (2l+1)=6$ is the dimension of the boson space and 
$m=\sum_j (2j+1)$ of the fermion space \cite{susy}. In this framework, 
even-even and odd-mass nuclei form the members of a supermultiplet. The 
inclusion of the neutron-proton degree of freedom leads to supersymmetric 
quartets of nuclei consisting of an even-even, an odd-even, an even-odd 
and an odd-odd nucleus \cite{quartet}. 

Supersymmetry (SUSY) distinguishes itself from other symmetries in that it 
includes, in addition to transformations among fermions or among bosons, 
also transformations between bosons and fermions. The spectroscopic 
properties of the nuclei that belong to the same supermultiplet, 
are linked and correlated by SUSY, i.e. they are described by the same form 
of the operators. Most tests of supersymmetry that have been discussed in the 
literature involve energies and transitions. These observables are described 
by the bosonic generators that transform bosons into bosons and fermions into 
fermions. Whereas the bosonic generators describe observables within a given 
nucleus, the fermionic generators that change a boson into a fermion or vice 
versa, describe the transitions between different nuclei of the same 
supermultiplet, such as observed in single-particle transfer reactions. 
Unlike for the bosonic sector, there are relatively few direct tests of the 
fermionic generators \cite{spin6,BI}. 

It is the purpose of this contribution to investigate one-nucleon transfer 
reactions in the context of nuclear supersymmetry, and to establish possible 
correlations between different transfer reactions. As an example, we consider 
the supersymmetric quartet of nuclei: $^{194}$Pt, $^{195}$Pt, $^{195}$Au 
and $^{196}$Au, whose energy spectra have been classified and described 
successfully in terms of the $U(6/12)_{\nu} \otimes U(6/4)_{\pi}$ 
supersymmetry \cite{quartet,metz}.

\section{The $U(6/12)_{\nu} \otimes U(6/4)_{\pi}$ supersymmetry}

The mass region $A \sim 190$ has been a rich source of possible empirical 
evidence for the existence of (super)symmetries in nuclei. The even-even 
nucleus $^{196}$Pt is the standard example of the $SO(6)$ limit of the 
IBM \cite{so6}. The odd-proton nuclei $^{191,193}$Ir and $^{193,195}$Au 
were suggested as examples of the $Spin(6)$ limit \cite{FI}, in which 
the odd-proton is allowed to occupy the $\pi d_{3/2}$ orbit, whereas the 
pairs of nuclei $^{192}$Os - $^{191}$Ir, $^{194}$Os - $^{193}$Ir, 
$^{192}$Pt - $^{193}$Au and $^{194}$Pt - $^{195}$Au have been analyzed as 
examples of a $U(6/4)$ supersymmetry \cite{susy}. The odd-neutron nucleus 
$^{195}$Pt, together with $^{194}$Pt, were studied in terms of a $U(6/12)$ 
supersymmetry, in which the odd neutron occupies the $\nu p_{1/2}$, 
$\nu p_{3/2}$ and $\nu f_{5/2}$ orbits \cite{baha}. 
These ideas were later extended to the case where neutron and proton bosons 
are distinguished \cite{quartet}, predicting in this way a correlation among 
quartets of nuclei, consisting of an even-even, an odd-proton, an odd-neutron 
and an odd-odd nucleus. The best experimental example of such a 
quartet with $U(6/12)_{\nu} \otimes U(6/4)_{\pi}$ supersymmetry is provided 
by the nuclei $^{194}$Pt, $^{195}$Au, $^{195}$Pt and $^{196}$Au. The number 
of bosons $N$ is taken to be the number of bosons in the odd-odd nucleus 
$^{196}$Au: $N=N_{\nu}+N_{\pi}=4+1=5$. 

\begin{table}[h]
\centering
\bea
\begin{array}{ccccccc}
\mbox{odd-odd} & & & & & & \mbox{even-odd} \\
N_{\nu}, N_{\pi}, j_{\nu}, j_{\pi} & & & & & & N_{\nu}+1, N_{\pi}, j_{\pi} \\
& & & & & & \\
& & ^{196}_{ 79}\mbox{Au}_{117} & \hspace{1cm} \leftrightarrow \hspace{1cm} 
& ^{195}_{ 79}\mbox{Au}_{116} & & \\
& & & & & & \\
& & \updownarrow & & \updownarrow & & \\
& & & & & & \\
& & ^{195}_{ 78}\mbox{Pt}_{117} & \leftrightarrow 
& ^{194}_{ 78}\mbox{Pt}_{116} & & \\
& & & & & & \\
\mbox{odd-even} & & & & & & \mbox{even-even} \\
N_{\nu}, N_{\pi}+1, j_{\nu} & & & & & & N_{\nu}+1, N_{\pi}+1 
\end{array}
\label{magic}
\eea
\end{table}

The supersymmetric classification of nuclear levels in the Pt 
and Au isotopes has been re-examined by taking advantage of the significant
improvements in experimental capabilities developed in the last decade.
High resolution transfer experiments with protons and polarized deuterons 
have led to strong evidence for the existence of supersymmetry (SUSY) in 
atomic nuclei. The experiments include high resolution transfer experiments 
to $^{196}$Au at TU/LMU M\"unchen \cite{metz,pt195}, and in-beam gamma ray 
and conversion electron spectroscopy following the reactions 
$^{196}$Pt$(d,2n)$ and $^{196}$Pt$(p,n)$ at the cyclotrons of the PSI and 
Bonn \cite{au196}. These studies have achieved an improved classification 
of states in $^{195}$Pt and $^{196}$Au which give further support to the 
original ideas \cite{baha,sun,quartet} and extend and refine previous 
experimental work \cite{mauthofer,jolie,rotbard} in this research area.

In a dynamical (super)symmetry, the Hamiltonian is expressed in terms of the 
Casimir invariants of the subgroups in a group chain. The relevant 
subgroup chain of $U(6/12)_{\nu} \otimes U(6/4)_{\pi}$ for the Pt and Au 
nuclei is given by \cite{quartet} 
\bea
U(6/12)_{\nu} \otimes U(6/4)_{\pi} &\supset& 
U^{B_{\nu}}(6) \otimes U^{F_{\nu}}(12) \otimes 
U^{B_{\pi}}(6) \otimes U^{F_{\pi}}(4) 
\nonumber\\
&\supset& U^B(6) \otimes U^{F_{\nu}}(6) \otimes U^{F_{\nu}}(2) \otimes 
U^{F_{\pi}}(4) 
\nonumber\\
&\supset& U^{BF_{\nu}}(6) \otimes U^{F_{\nu}}(2) \otimes U^{F_{\pi}}(4)
\nonumber\\
&\supset& SO^{BF_{\nu}}(6) \otimes U^{F_{\nu}}(2) \otimes SU^{F_{\pi}}(4)
\nonumber\\
&\supset& Spin(6) \otimes U^{F_{\nu}}(2) 
\nonumber\\
&\supset& Spin(5) \otimes U^{F_{\nu}}(2) 
\nonumber\\
&\supset& Spin(3) \otimes SU^{F_{\nu}}(2) 
\nonumber\\
&\supset& SU(2) ~.
\eea
In this case, the Hamiltonian
\bea
H &=& \alpha \, C_{2U^{BF_{\nu}}(6)} + \beta \, C_{2SO^{BF_{\nu}}(6)} 
+ \gamma \, C_{2Spin(6)} 
\nonumber\\
&& + \delta \, C_{2Spin(5)} + \epsilon \, C_{2Spin(3)} 
+ \eta \, C_{2SU(2)} ~, 
\eea
describes simultaneously the excitation spectra of the quartet of nuclei. 
Here we have neglected terms that only contribute to binding energies. 
The coefficients $\alpha$, $\beta$, $\gamma$, $\delta$, $\epsilon$ and 
$\eta$ have been determined in a simultaneous fit of the excitation energies 
of the four nuclei of Eq.~(\ref{magic}) \cite{au196}. 

In a dynamical supersymmetry, closed expressions can be derived for energies, 
and selection rules and intensities for electromagnetic transitions and 
single-particle transfer reactions. While a simultaneous description and 
classification of these observables in terms of the 
$U(6/12)_{\nu} \otimes U(6/4)_{\pi}$ supersymmetry has been shown to 
be fulfilled to a good approximation for the quartet of nuclei $^{194}$Pt, 
$^{195}$Au, $^{195}$Pt and $^{196}$Au, there are important predictions 
still not fully verified by experiments. These tests involve the transfer 
reaction intensities among the supersymmetric partners. In the next section 
we concentrate on the latter and, in particular, on the one-proton transfer 
reactions $^{194}$Pt $\rightarrow$ $^{195}$Au and 
$^{195}$Pt $\rightarrow$ $^{196}$Au. 

\section{One-nucleon transfer reactions}

The single-particle transfer operator that is commonly used in the 
Interacting Boson-Fermion Model (IBFM), has been derived in the seniority 
scheme \cite{olaf}. Although strictly speaking this derivation is only valid 
in the vibrational regime, it has been used for deformed nuclei as well. 
An alternative method is based on symmetry considerations. It consists in 
expressing the single-particle transfer operator in terms of tensor operators 
under the subgroups that appear in the group chain of a dynamical 
(super)symmetry \cite{spin6,BI,barea}. The single-particle transfer between 
different members of the same supermultiplet provides an important test of 
supersymmetries, since it involves the transformation of a boson into a 
fermion or vice versa, but it conserves the total number of bosons plus 
fermions. 

The operators that describe one-proton transfer reactions in the 
$U(6/12)_{\nu} \otimes U(6/4)_{\pi}$ supersymmetry are given by 
\bea
T_{1,m}^{(\frac{1}{2},\frac{1}{2},-\frac{1}{2}),(\frac{1}{2},\frac{1}{2}),
\frac{3}{2}} &=& -\sqrt{\frac{1}{6}} \left( \tilde{s}_{\pi} \times 
a^{\dagger}_{\pi,\frac{3}{2}} \right)^{(\frac{3}{2})}_m 
+\sqrt{\frac{5}{6}} \left( \tilde{d}_{\pi} \times 
a^{\dagger}_{\pi,\frac{3}{2}} \right)^{(\frac{3}{2})}_m ~, 
\nonumber\\ 
T_{2,m}^{(\frac{3}{2},\frac{1}{2},\frac{1}{2}),(\frac{1}{2},\frac{1}{2}),
\frac{3}{2}} &=& \sqrt{\frac{5}{6}} \left( \tilde{s}_{\pi} \times 
a^{\dagger}_{\pi,\frac{3}{2}} \right)^{(\frac{3}{2})}_m 
+\sqrt{\frac{1}{6}} \left( \tilde{d}_{\pi} \times 
a^{\dagger}_{\pi,\frac{3}{2}} \right)^{(\frac{3}{2})}_m ~.
\label{top1} 
\eea
The operators $T_1$ and $T_2$ are, by construction, tensor operators under 
$Spin(6)$, $Spin(5)$ and $Spin(3)$ \cite{barea}. The upper indices 
$(\sigma_{1},\sigma_{2},\sigma_{3})$, $(\tau_{1},\tau_{2})$, $J$ 
specify the tensorial properties under $Spin(6)$, $Spin(5)$ and $Spin(3)$.
The use of tensor operators to describe single-particle transfer 
reactions in the supersymmetry scheme has the advantage of giving rise to 
selection rules and closed expressions for the spectroscopic 
factors. 

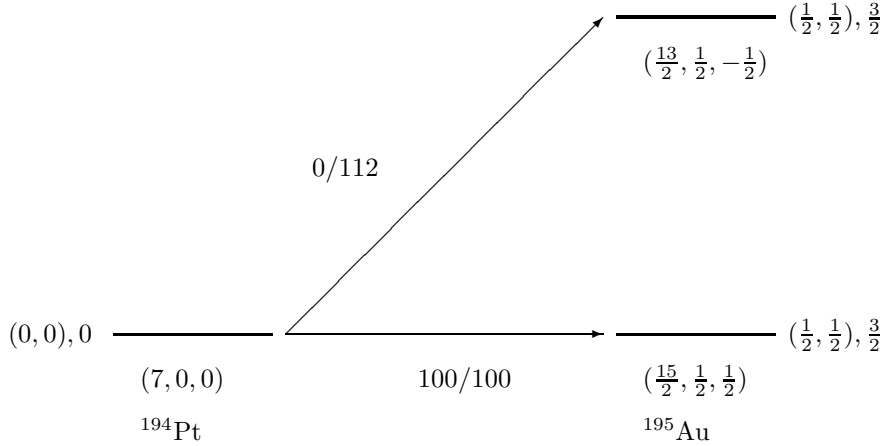
\begin{figure}
\centering
\setlength{\unitlength}{1.0pt}
\begin{picture}(400,200)(0,0)
\thicklines
\put ( 50, 60) {\line(1,0){60}}
\put (240, 60) {\line(1,0){60}}
\put (240,180) {\line(1,0){60}}
\put (125,120) {$0/112$}
\put (165, 40) {$100/100$}
\put ( 60, 40) {$(7,0,0)$}
\put ( 60, 20) {$^{194}$Pt}
\put ( 10, 57) {$(0,0),0$}
\put (250, 40) {$(\frac{15}{2},\frac{1}{2},\frac{1}{2})$}
\put (250, 20) {$^{195}$Au}
\put (305, 57) {$(\frac{1}{2},\frac{1}{2}),\frac{3}{2}$}
\put (250,160) {$(\frac{13}{2},\frac{1}{2},-\frac{1}{2})$}
\put (305,177) {$(\frac{1}{2},\frac{1}{2}),\frac{3}{2}$}
\thinlines
\put (115, 60) {\vector( 1, 0){120}}
\put (115, 60) {\vector( 1, 1){120}}
\end{picture}
\caption[]{Allowed one-proton transfer reactions for 
$^{194}$Pt $\rightarrow$ $^{195}$Au. The spectroscopic factors 
are normalized to 100 for the ground state to ground state 
transition for the operators $T_1/T_2$.}
\label{spec1}
\end{figure}

Fig.~\ref{spec1} shows the allowed transitions for the transfer operators 
of Eq.~(\ref{top1}) that describe the one-proton transfer from the ground 
state $|(N+2,0,0),(0,0),0\rangle$ of the even-even nucleus $^{194}$Pt to 
the even-odd nucleus $^{195}$Au belonging to the same supermultiplet 
$[N_{\nu}+1\}_{\nu} \otimes [N_{\pi}+1\}_{\pi}$. The operators $T_1$ and 
$T_2$ have the same transformation character under $Spin(5)$ and $Spin(3)$, 
and therefore can only excite states with 
$(\tau_1,\tau_2)=(\frac{1}{2},\frac{1}{2})$ and $J=\frac{3}{2}$. However, 
they differ in their $Spin(6)$ selection rules. Whereas $T_1$ 
can only excite the ground state of the even-odd nucleus with 
$(\sigma_1,\sigma_2,\sigma_3)=(N+\frac{3}{2},\frac{1}{2},\frac{1}{2})$, 
the operator $T_2$ also allows the transfer to an excited state with 
$(N+\frac{1}{2},\frac{1}{2},-\frac{1}{2})$. 
The ratio of the intensities is given by \cite{barea} 
\bea
R_1 &=& \frac{I_{\rm gs \rightarrow exc}}{I_{\rm gs \rightarrow gs}} 
\;=\; 0 ~,
\nonumber\\
R_2 &=& \frac{I_{\rm gs \rightarrow exc}}{I_{\rm gs \rightarrow gs}} 
\;=\; \frac{9(N+1)(N+5)}{4(N+6)^2} ~, 
\label{ratios}
\eea
for $T_1$ and $T_2$, respectively. In the case of the one-proton transfer 
$^{194}$Pt $\rightarrow$ $^{195}$Au, the second ratio is given by 
$R_2=1.12$ ($N=5$). 

The available experimental data from the proton stripping reactions 
$^{194}$Pt$(\alpha,t)^{195}$Au and $^{194}$Pt$(^{3}$He$,d)^{195}$Au 
\cite{munger} shows that the $J=3/2$ ground state of $^{195}$Au is excited 
strongly with $C^2S=0.175$, whereas the first excited $J=3/2$ state is 
excited weakly with $C^2S=0.019$. In the SUSY scheme, the latter state is 
assigned as a member of the ground state band with 
$(\tau_1,\tau_2)=(5/2,1/2)$. Therefore the one proton transfer to this 
state is forbidden by the $Spin(5)$ selection rule of the tensor operators 
of Eq.~(\ref{top1}). The relatively small strength to excited $J=3/2$ 
states suggests that the operator $T_1$ of Eq.~(\ref{top1}) can be used 
to describe the data. 

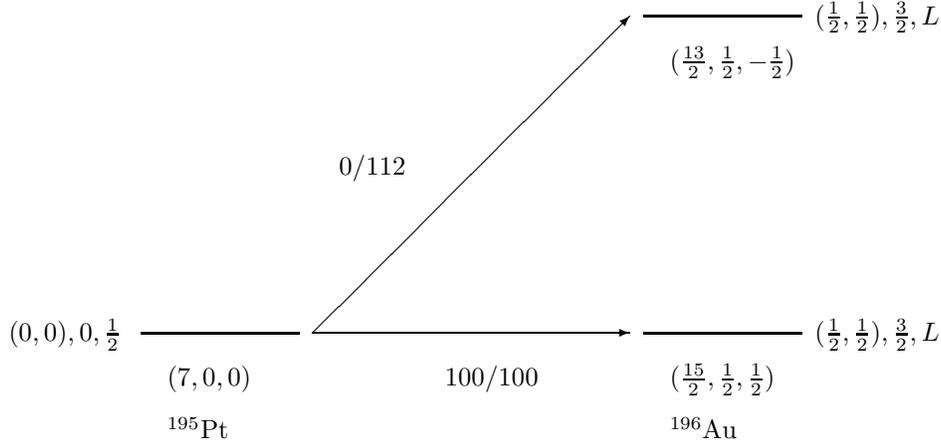
\begin{figure}
\centering
\setlength{\unitlength}{1.0pt}
\begin{picture}(400,200)(0,0)
\thicklines
\put ( 50, 60) {\line(1,0){60}}
\put (240, 60) {\line(1,0){60}}
\put (240,180) {\line(1,0){60}}
\put (125,120) {$0/112$}
\put (165, 40) {$100/100$}
\put ( 60, 40) {$(7,0,0)$}
\put ( 60, 20) {$^{195}$Pt}
\put (  0, 57) {$(0,0),0,\frac{1}{2}$}
\put (250, 40) {$(\frac{15}{2},\frac{1}{2},\frac{1}{2})$}
\put (250, 20) {$^{196}$Au}
\put (305, 57) {$(\frac{1}{2},\frac{1}{2}),\frac{3}{2},L$}
\put (250,160) {$(\frac{13}{2},\frac{1}{2},-\frac{1}{2})$}
\put (305,177) {$(\frac{1}{2},\frac{1}{2}),\frac{3}{2},L$}
\thinlines
\put (115, 60) {\vector(1,0){120}}
\put (115, 60) {\vector(1,1){120}}
\end{picture}
\caption[]{As Fig.~\ref{spec1}, but for 
$^{195}$Pt $\rightarrow$ $^{196}$Au.}
\label{spec2}
\end{figure}

In Fig.~\ref{spec2} we show the allowed transitions for the one-proton 
transfer from the ground state $|(N+2,0,0),(0,0),0,\frac{1}{2}\rangle$ 
of the odd-even nucleus $^{195}$Pt to the odd-odd nucleus $^{196}$Au. 
Also in this case, the operator $T_1$ only excites the ground state doublet 
of $^{196}$Au with 
$(\sigma_1,\sigma_2,\sigma_3)=(N+\frac{3}{2},\frac{1}{2},\frac{1}{2})$, 
$(\tau_1,\tau_2)=(\frac{1}{2},\frac{1}{2})$, $J=\frac{3}{2}$ and 
$L=J\pm\frac{1}{2}$, whereas $T_2$ also populates the excited state with 
$(N+\frac{1}{2},\frac{1}{2},-\frac{1}{2})$. The ratio of the intensities 
is the same as for the $^{194}$Pt $\rightarrow$ $^{195}$Au transfer 
reaction 
\bea
R_1(^{195}\mbox{Pt} \rightarrow ^{196}\mbox{Au}) &=& 
R_1(^{194}\mbox{Pt} \rightarrow ^{195}\mbox{Au}) \;=\; 0 ~,
\nonumber\\
R_2(^{195}\mbox{Pt} \rightarrow ^{196}\mbox{Au}) &=& 
R_2(^{194}\mbox{Pt} \rightarrow ^{195}\mbox{Au}) \;=\; 
\frac{9(N+1)(N+5)}{4(N+6)^2} ~.
\eea
This is direct consequence of the supersymmetry. Just as the energies 
and the electromagnetic transition rates of the supersymmetric quartet of 
nuclei were calculated with the same form of the Hamiltonian and the 
transition operator, here we have extended this idea to the one-proton 
transfer reactions. We find definite predictions for the spectroscopic 
factors of the $^{195}$Pt $\rightarrow$ $^{196}$Au transfer reactions, 
which can be tested experimentally. To the best of our knowledge, there 
are no data available for this reaction. 

For the one-neutron transfer reactions there exists a similar situation. 
The available experimental data from the neutron stripping reactions 
$^{194}$Pt$(d,p)^{195}$Pt \cite{sheline} can be used to determine the 
appropriate form of the one-neutron transfer operator \cite{BI}, which 
then can be used to predict the spectroscopic factors for the transfer 
reaction $^{195}$Au $\rightarrow$ $^{196}$Au. We believe that, as a 
consequence of the supersymmetry classification, a number of additional 
correlations exist for transfer reactions between different pairs of nuclei. 
This would be the first time that such relations are predicted for nuclear 
reactions which may provide a challenge and motivation for future 
experiments. 

\section{Summary and outlook}

The recent measurements of the spectroscopic properties of the odd-odd 
nucleus $^{196}$Au have rekindled the interest in nuclear supersymmetry. 
The available data on the spectroscopy of the quartet of nuclei $^{194}$Pt, 
$^{195}$Au, $^{195}$Pt and $^{196}$Au can, to a good approximation, be 
described in terms of the $U(6/4)_{\pi} \otimes U(6/12)_{\nu}$ 
supersymmetry. However, there is a still another important set of experiments 
which can further test the predictions of the supersymmetry scheme. 
These involve transfer reactions between nuclei belonging to the same 
supermultiplet, in particular between the even-odd (odd-even) and odd-odd 
members of the supersymmetric quartet. Theoretically, these transfers are 
described by the supersymmetric generators which change a boson into a 
fermion, or vice versa. 

We investigated in some detail the example of proton transfer between the 
SUSY partners: $^{194}$Pt $\rightarrow$ $^{195}$Au and 
$^{195}$Pt $\rightarrow$ $^{196}$Au. The supersymmetry implies strong 
correlations for the spectroscopic factors of these two reactions which 
can be tested experimentally. A similar set of relations can be derived for 
the one-neutron transfer reactions $^{194}$Pt $\leftrightarrow$ $^{195}$Pt 
and $^{195}$Au $\leftrightarrow$ $^{196}$Au. Another interesting extension 
of supersymmetry concerns the recently measured two-nucleon transfer reaction 
$^{194}$Pt$(\alpha,d)^{196}$Au \cite{graw}, in which a neutron-proton pair 
is transferred to the target nucleus. This reaction presents a very sensitive 
test of the wave functions, since it provides a measure of the correlation 
within the transferred neutron-proton pair. Whether it is possible to describe 
this process by a transfer operator that is correlated by SUSY to that 
of the one-proton and one-neutron transfer reactions is an open question. 

In conclusion, we emphasize the need for new experiments taking advantage  
of the new experimental capabilities \cite{metz,pt195,au196} and suggest 
that particular attention be paid to one- and two-nucleon transfer reactions 
between the SUSY partners $^{194}$Pt, $^{195}$Au, $^{195}$Pt and $^{196}$Au, 
since such experiments provide the most stringent tests of nuclear 
supersymmetry. It remains to be seen whether the correlations predicted by 
SUSY are indeed verified by experiments. 

\section*{Dedication}

It is a great pleasure to dedicate this contribution to Franco Iachello 
on the occasion of the conference `Symmetries in Nuclear Structure', held 
in his honor. Unfortunately it was not possible to attend the meeting, but 
I am grateful to the organizers for the invitation to write a 
contribution for the proceedings. 

In the fall of 1977, I took a course on nuclear structure presented by a 
young Italian professor at the University of Groningen. The lectures were 
characterized by their clarity of presentation, a contagious enthusiasm, a 
link between the material presented in the course and ongoing research and, 
last but not least, the connection between theory and experiment. These 
ingredients have formed the basis and provided the motivation and inspiration 
for my own scientific career, first as a graduate student at the KVI in 
Groningen, and later as a research scientist. Over the years I have had the 
pleasure to collaborate with Franco on different subjects, such as 
supersymmetry, baryon spectroscopy and nuclear clusters which has resulted 
in 12 joint publications between 1983 and 2002. 

I will not mention his other career as a racecar driver, nor comment on his 
uncanny likeliness to Woody Allen (with the exception of the title of this 
contribution). Finally, I wish Franco an equally productive and creative 
second half of his career. Congratulations, Franco! 

\section*{Acknowledgments}

We are grateful to Gerhard Graw for numerous stimulating discussions 
and for sharing the new data on the two-nucleon transfer reactions 
$^{198}$Hg$(\vec{d},\alpha)^{196}$Au and $^{194}$Pt$(\alpha,d)^{196}$Au 
prior to publication. The work presented in this contribution is motivated 
by the renewed experimental interest in this mass region. 
This work was supported in part by CONACyT.

\end{document}